# Thickness-Driven Superconductor-Insulator Transition in (Cu,C)-1234 and Proximity-Induced Superconductivity Recovery in (Cu,C)-1234/YBCO Heterostructure

Zhihang Xu, Detian Yang, Ping Zhu, Ruoxian Sun, Xiaoyang Cai, Yanqun Guo*, Chuanbing Cai*

Shanghai Key Laboratory of High Temperature Superconductors, Department of Physics, Shanghai University, Shanghai 200444, People's Republic of China

**Abstract**

Superconducting proximity effect and related thickness-driven property evolution remain an important issue in understanding high temperature superconductors. Among proximity systems, superconductor-superconductor (S-S') is special for the existence of intrinsic superconductivity in both materials. Such platform allows the different superconducting orders to compete, couple and reconstruct at the interface. In this paper, (Cu,C)-1234/YBCO heterostructure grown on LAO (001) with fixed thickness of bottom YBCO layer as 150 nm and varied thickness of top (Cu,C)-1234 layer as 188nm, 87 nm, 18nm and estimated 1.2 nm were fabricated and component films were preserved. Electrical transport characterization indicated that as the thickness decrease the (Cu,C)-1234 film degrades and underwent the superconductor-insulator transition (SIT) from thicker to less than 18 nm. In contrast, superconductivity is re-established in transport measurements when the insulating (Cu,C)-1234 layer is coupled to superconducting YBCO As the (Cu,C)-1234 thickness is further reduced to approximately 1.2 nm, the recovered superconductivity is strongly suppressed. The observed thickness dependence is consistent with a scenario in which interfacial coupling restores superconductivity over a finite thickness range before increasing disorder and dimensional confinement dominate in the two-dimensional limit. This work establishes a promising platform for investigating interfacial coupling between cuprate superconductors and provides new insight into the superconducting proximity effect in high-temperature superconducting heterostructures.

## 1. Introduction

Superconducting proximity effect is one of the fundamental phenomena in superconductivity and provides an effective approach for engineering quantum states in hybrid materials. It has been extensively investigated in systems including superconductor-antiferromagnet (S-AF) [1], superconductor-ferromagnet (S-F), superconductor-normal metal (S-N), superconductor-topological insulator (S-TI) and more complicated multilayer structures, where interesting quantum phenomena like Andreev reflection, Majorana fermion, mixed pairing, topological superconductivity emerge. These studies have established the proximity effect as a powerful tool for exploring novel quantum phenomena and developing superconducting electronic devices.

In different systems, superconducting proximity effect functioned in different aspects. For S-N, Andreev reflection and Cooper pairs leakage from S to N is the main process which induce superconductivity in normal metal. For S-F, superconducting order and ferromagnetic order compete with each other at interface leading to properties leaked to the other and unconventional superconducting pairing. For S-AF, Néel order leaks into superconductor causing Néel triplets which breaks original spin singlets in superconductor so that lower the $T_c$. For S-TI, the surface state of topological insulator couples with the superconductor to induce topological superconductivity at the surface as a novel state which supports the emergence of Majorana fermions. Further applications like quantum calculation and spintronic devices were under development based on these findings. Compared with these systems, in superconductor-superconductor(S-S') heterostructures two different superconducting orders could couple, compete or reconstruct [11-13]. Instead of simply inducing superconductivity in a nonsuperconducting material, the interface involves the mutual interaction and reconstruction of two superconducting condensates. Their coupling is influenced by the superconducting gap, coherence length, pairing symmetry, and electronic correlations of each material, leading to phenomena that cannot be fully explained by conventional proximity theories. The mismatch of pairing symmetry and potential coupling at interface finally gives rise to mixed pairing amplitude, crossed Andreev reflection and new pairing mechanism which make S-S' an irreplaceable platform to explore the microscopic mystery of high temperature superconductor.

Besides interface coupling, dimensionality is another key parameter governing superconductivity in thin films. In cuprate superconductors, however, the exact influence is still controversial. For Bi-2212 [14], the property of single layer is the same as the bulk sample indicating that high temperature superconductivity is able to exist even in single layer limit which defies the expectation from Mermin-Wagner theorem. However, for LSCO [15] the $T_c$ decreases when the thickness is less than four unit cells and vanishes when the thickness is less than two unit cells because of the weaken of interlayer coupling. For YBCO with $T_c$=90.5 K[16], by high-resolution torque-magnetometry it was shown that at 80 K above the data is consistent with three-dimensional phenomenological theory but an anomalous torque develops when below 80 K which provide the evidence of the transition from three-dimensional superconducting behavior to two-dimensional superconducting behavior. For S-TI heterostructure $Bi_2Te_3$/Bi-2212 [10], as the thickness of $Bi_2Te_3$ increase, the maximum of superconducting gap decays exponentially indicating the enhancement of uncoupling. And for F-S-F sandwich LCMO-YBCO-LCMO [19], when the thickness of YBCO decreases to five unit cells, the superconductivity was suppressed by the penetration of antiferromagnetic coupling and proposed inverse spin-switch effect as the explanation for the anomalous increase of $T_c$ as the magnetic field increase. It is necessary to set thickness as a crucial variable in heterostructures with high temperature superconductors to get in-depth understanding of the interaction between superconductor and other materials.

SIT [15-17]has attracted considerable attention because it represents a quantum phase transition between two fundamentally different electronic ground states. Although its microscopic mechanism remains under active debate, increasing experimental evidence suggests that the transition is governed primarily by enhanced disorder and reduced phase stiffness rather than by the complete disappearance of Cooper pairing. In this picture, superconducting pairs may remain locally intact after long-range phase coherence has been destroyed, giving rise to an insulating state containing localized Cooper pairs. Transport studies on epitaxial $La_{2-x}Sr_xCuO_4$ thin films have demonstrated a clear thickness-controlled SIT near a critical thickness of approximately 10 nm, accompanied by resistance upturns indicative of weak localization and enhanced phase fluctuations [15]. These results indicate that dimensional confinement primarily suppresses global phase coherence while superconducting pairing may partially survive beyond the transition.

Despite significant progress in both superconducting proximity effects and thickness-controlled superconductivity, these two research directions have largely been investigated independently. Previous studies on proximity effects have mainly focused on inducing superconductivity in nonsuperconducting materials or understanding interfacial superconducting coupling, whereas investigations of thickness-driven SIT have almost exclusively examined isolated superconducting films. Consequently, how does a thickness-driven superconductor–insulator transition influence the superconducting proximity effect between two intrinsic high-temperature superconductors remains unanswered. If superconducting pairing survives locally after long-range phase coherence is lost, coupling to a neighboring superconductor may restore phase coherence through the proximity effect and recover superconductivity. Understanding this interplay is essential for clarifying the relationship among dimensionality, disorder, phase coherence, and interfacial superconductivity in strongly correlated superconductors.

In this work, (Cu,C)-1234/YBCO heterostructure constituted by two cuprate superconducting films was fabricated. By setting thickness of (Cu,C)-1234 as the variable and sole thin film reserved as references, unexpected superconductor-insulator transition emerged in ultrathin (Cu,C)-1234 and abnormal proximity effect appeared in circuit going through the interface. The results of electrical transport of sole (Cu,C)-1234 shown that decreased-thickness-induced disorder enhancement and consequently strong localization of Cooper pairs in (Cu,C)-1234 caused the loss of phase coherence which broke superconductivity. In heterostructures, extremely thin (Cu,C)-1234 which was supposed to be insulating regained its superconductivity by proximity effect. The phase coherence in (Cu,C)-1234 was reestablished because of phase reference from robust superconductor YBCO. The sample with final thin 1.2 nm (Cu,C)-1234 got degraded superconducting properties which indicated that thickness comparable to coherent length had already also broke pairing to some extent besides the enhancement of disorder and related localization that destroy long-range coherence.

## 2. Methods

### 2.1 Target preparation

The YBCO target was provided by Hefei Kejing Material Technology Co., LTD, and the $Ba_2Ca_3Cu_{4.6}O_y$ target was made by solid reaction method. The precursor powders including $BaCO_3$, $CaCO_3$ and CuO with purities over 99.5% were used to synthesis the target with nominal composition $Ba_2Ca_3Cu_{4.6}O_y$. After sufficient mixing and grinding, the mixed powder was then heated in sequential procedures at 800 °C and 840 °C for 24 h in air with intermediate grinding.

Then the mixed powder was ground and pressed into a tablet, a hydrostatic pressure of 180 MPa was exerted to stabilize the target. Finally, the tablet was sintered at 880 °C for 48 h in air.

### 2.2 Thin film growth

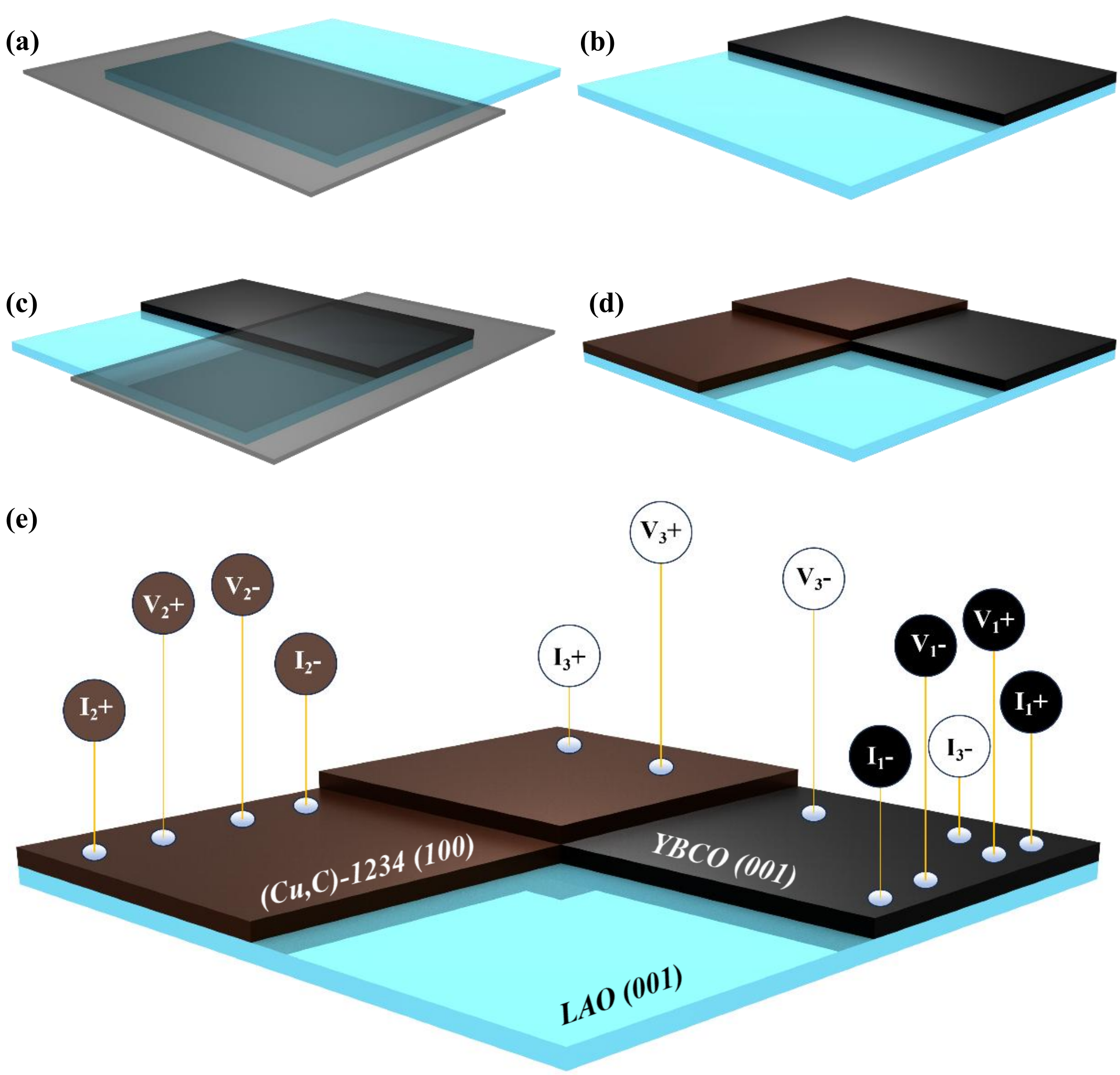

**Fig. 1.** The growth procedure of the heterostructure and three test circuits to obtain the temperature-dependent resistance of single thin films and interface by four-probe method. (a) LAO substrate with the surface half covered by thin wafer. (b) YBCO deposited at half the substrate. (c) both grown YBCO and LAO substrate half covered by the wafer. (d) the final four-grid area distribution including LAO substrate, sole YBCO, sole (Cu,C)-1234 and (Cu,C)-1234/YBCO heterostructure after the deposition of (Cu,C)-1234. (e) Test circuits for sole YBCO, sole (Cu,C)-1234 and interface.

A KrF excimer laser (λ = 248 nm, COHERENT Compex Pro 205F) was used to prepare (Cu,C)-1234/YBCO heterostructure by pulsed laser deposition method. To confirm the quality of component films of the heterostructure and make it possible to conduct the tests for each of them and the interface, the procedure of fabrication shown in Fig. 1 was taken. The LAO (001) substrate was glued to a stainless-steel sample table by silver paste. Before the deposition, a piece of thin wafer was added to cover nearly half the LAO (001) substrate. To make the wafer as close to the substrate as it can to protect the area covered, another substrate was glued under it to keep the wafer at the same height as the substrate for deposition. The substrate was put with 4.5 m distance from the target. Since the deposition temperature of YBCO is nearly 100℃ higher than (Cu,C)-1234 which can damage the latter, YBCO thin film was set as the first layer. For the deposition of YBCO, the parameters were set as 830 ℃ for deposition temperature, 10 min for growth time, 30 $cm^3$/min for the flow rate of $O_2$, 20 Pa for the chamber pressure during growth, 300 mJ/pulse and 5 Hz for the laser, 600 ℃ and 60 min for annealing and oxygen absorption after deposition. Then another wafer was added perpendicularly to replace the prior one and continue the deposition of (Cu,C)-1234 with the parameters as 740 ℃ for deposition temperature, 30 $cm^3$/min:20$cm^3$/min for the flow rate of $O_2$ and $CO_2$, 20 Pa for the chamber pressure during growth, 300 mJ/pulse and 5 Hz for the laser, 600 ℃ and 60 min for annealing and oxygen absorption after deposition. The deposition duration for (Cu,C)-1234 was respectively set as 10min, 5min, 1min, 4s for S1-S4.

### 2.3 Structural characterization

The crystal structures of the individual films and heterostructures were characterized by X-ray diffraction (XRD) using a Rigaku SmartLab diffractometer with a 2θ scanning range of 10°–60°. Film thicknesses and surface morphologies were examined using a field-emission scanning electron microscope (Hitachi Regulus 8230). Cross-sectional high-angle annular dark-field scanning transmission electron microscopy (HAADF-STEM) images were acquired using a JEOL JEM-ARM300F GRAND ARM microscope. Fast Fourier transform (FFT) analyses of the STEM images were performed using DigitalMicrograph software to evaluate the crystallographic relationship across the heterointerface.

### 2.4 Electrical transport measurements

Electrical transport measurements were performed in a Physical Property Measurement System (PPMS-9T, Quantum Design) using the standard four-probe method. Owing to the shadow-mask design, three independent measurement configurations were fabricated on the same substrate (figure 1), allowing the temperature-dependent resistance of the single YBCO film, the single (Cu,C)-1234 film, and the (Cu,C)-1234/YBCO heterostructure to be measured under identical experimental conditions. This configuration minimizes sample-to-sample variations and enables a direct evaluation of the superconducting proximity effect by comparing the transport behavior of the isolated films with that of the heterostructure.

## 3. Results

### 3.1 structural characterization

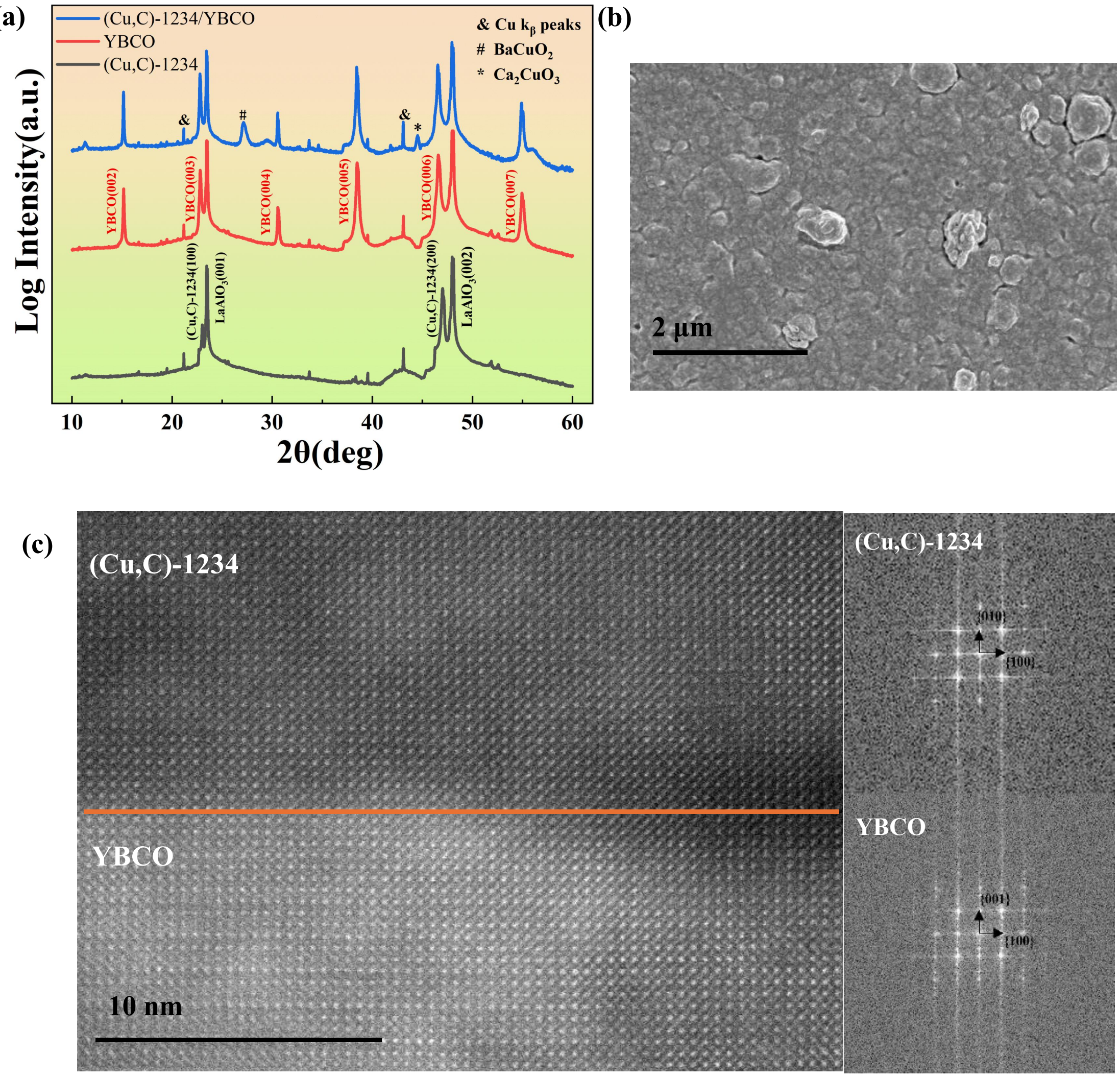

**Fig. 2.** (a) XRD pattern of (Cu,C)-1234 (black), YBCO (red) and heterostructure (blue) with highest $T_c$. (b) 10Kx SEM image of (Cu,C)-1234/YBCO heterostructure. (c) HAADF-STEM image and FFT results of (Cu,C)-1234 and YBCO.

XRD pattern of YBCO, (Cu,C)-1234 and heterostructure in log plot respectively was shown in Fig. 2. (a). Using the diffraction peaks of LAO (001) and LAO (002) as references, it is obvious to see that the single films both got phase in good quality. Though the diffraction peak of (Cu,C)-1234 (100) and (Cu,C)-1234 (200) is very close to YBCO (003) and YBCO (006), the minor existence of impurity phases of $BaCuO_2$ and $Ca_2CuO_3$ together with the slight increase of the log intensity of the peaks near LAO (001) and LAO (002) provided the evidence that the heterostructure with good quality of each layer was successfully fabricated. The surface morphology of (Cu,C)-1234/YBCO was shown in Fig. 2(b). The top (Cu,C)-1234 film possessed good continuity and uniformity with several particles. Using HAADF-STEM, atoms arrangement near the interface and FFT analyzation of the heterostructure were obtained illustrated in Fig. 2 (c). In YBCO, well-arranged c-axis oriented YBCO could be seen and the orientation was confirmed by the FFT analyze result (the distance calibrated were 0.3808 nm laterally and 1.1522 nm vertically which are respectively consistent with (100) and (001) interplanar distances for YBCO). In (Cu,C)-1234, the arrangement was able to be identified from the image and the orientation was verified by FFT in which the calibrated distances of both lateral and vertical directions are 0.3802 nm that are consistent with the interplanar distances of (Cu,C)-1234 (100) and (010) orientation.

### 3.2 Electronic transport measurements and thickness-driven SIT in (Cu,C)-1234

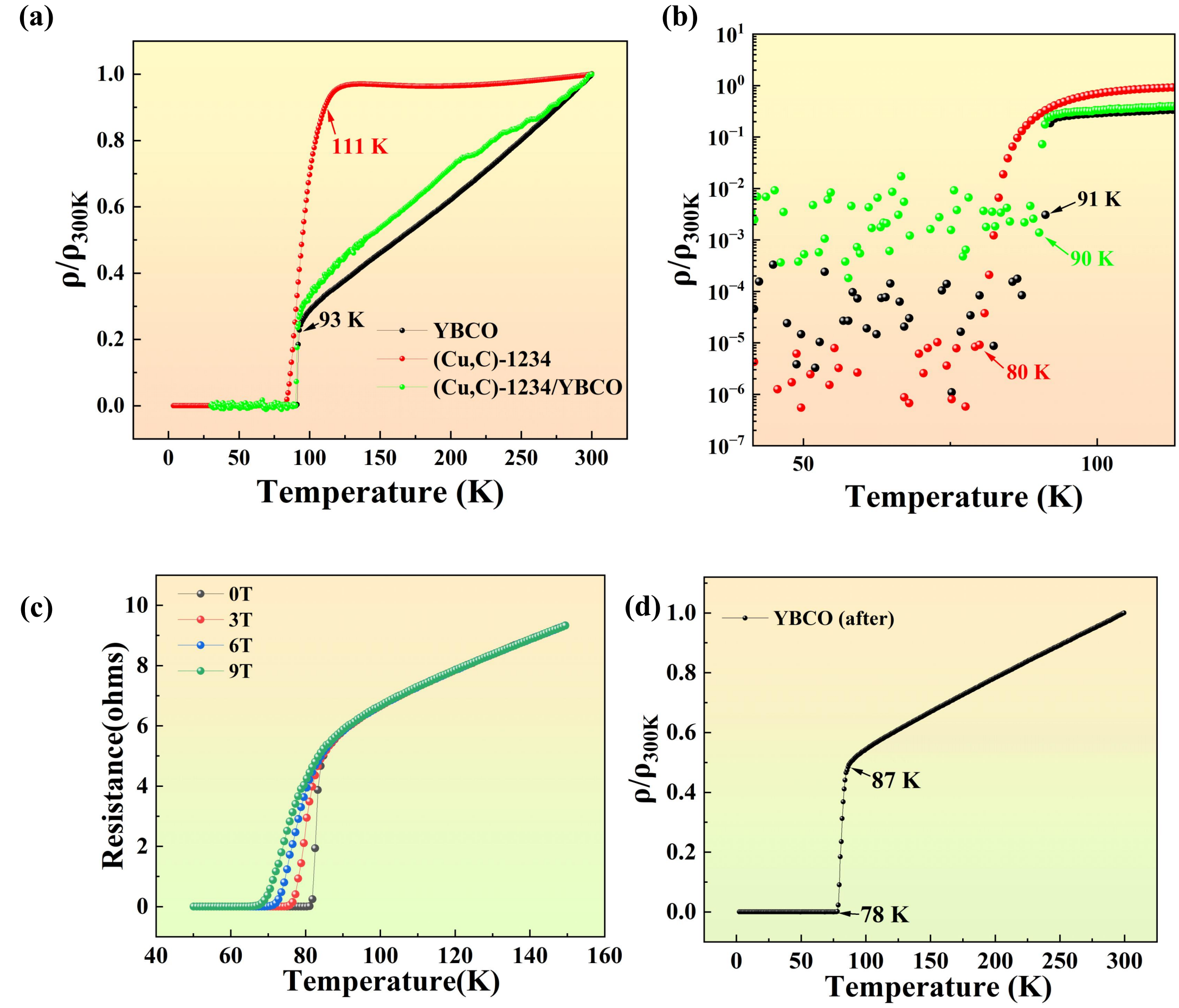

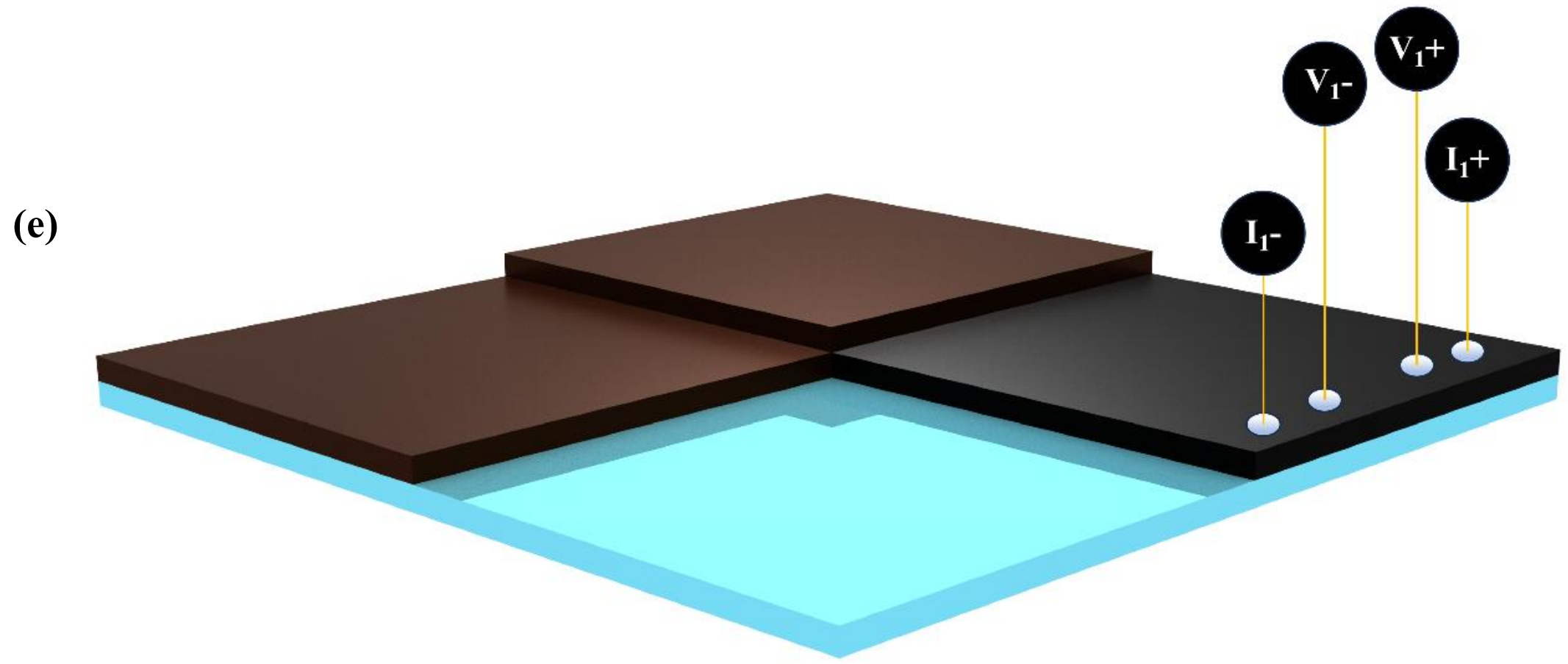

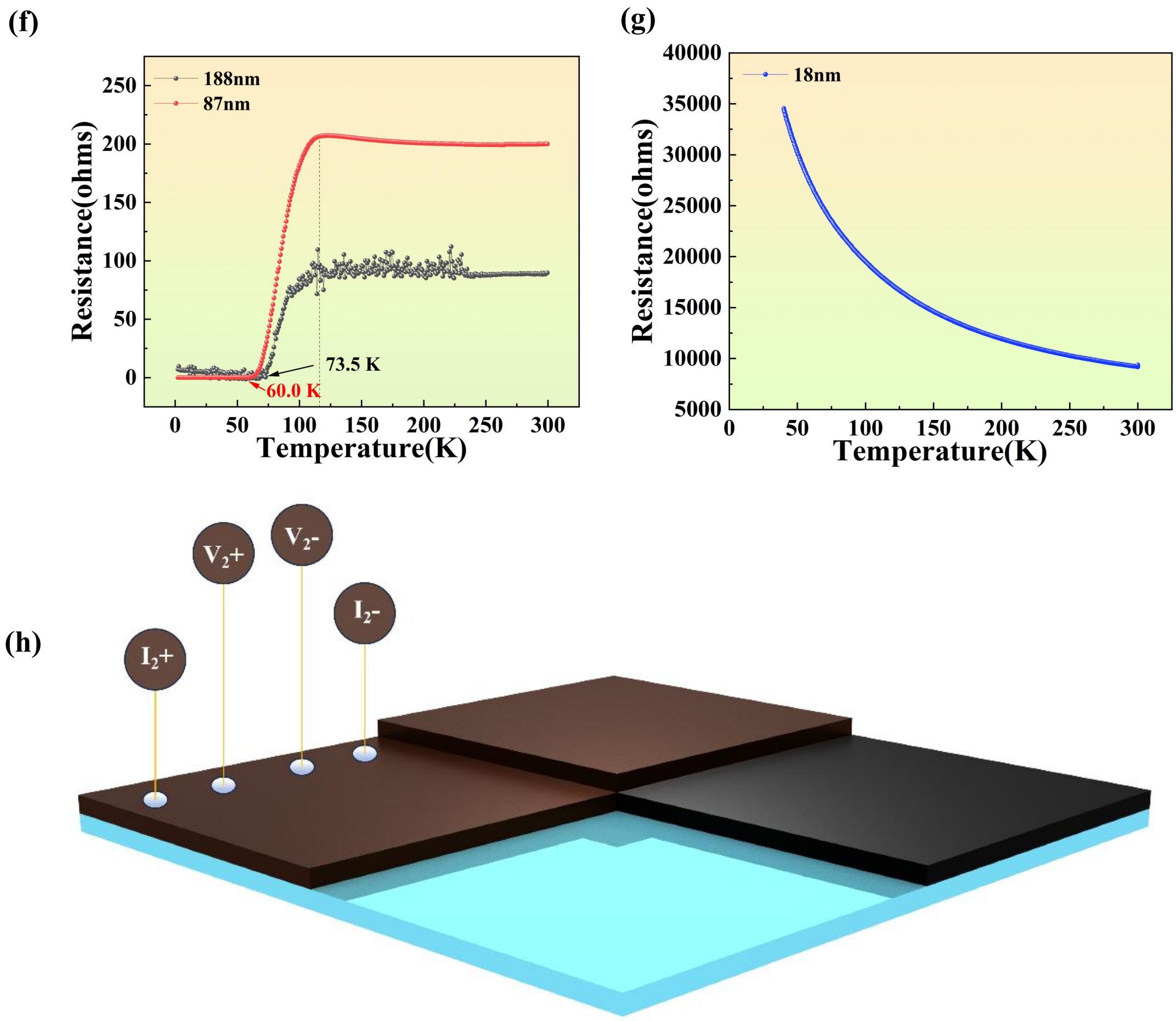

**Fig. 3.** (a) Normalized ρ-T curve of solely grown YBCO (black), solely grown (Cu,C)-1234 (red) and (Cu,C)-1234/YBCO heterostructure. Here the circuit for heterostructure is on the top (Cu,C)-1234 layer and the heterostructure was in-situ grown. (b) Enlarged view of $T_{c0}$ region in (a). (c) R-T curves for another heterostructure sample under 0-9 T (d) Normalized ρ-T curve of YBCO after the growth process of (Cu,C)-1234. The $T_{c0}$ dropped from 91K to 78 K. (e)Test circuit for YBCO. (f) R-T curve of (Cu,C)-1234 with thickness of 188 nm and 87 nm. (g) R-T curve of (Cu,C)-1234 with thickness of 18 nm which transited into insulator. (h) Test circuit for (Cu,C)-1234.

Fig. 3 (a) shows the temperature-dependent normalized ρ-T curve of two component thin films grown solely and (Cu,C)-1234/YBCO heterostructure. Here the circuit for heterostructure is on the top layer (Cu,C)-1234. Fig. 3 (b) is the enlarged view of region near $T_{c0}$ of each sample. The $T_{c0}$ is 80 K for (Cu,C)-1234, 91 K for YBCO and 90 K for the heterostructure. There is no significant difference between YBCO and heterostructure tested on top (Cu,C)-1234 layer. To avoid possible parallel conduction [20] and ensure the current flows through the interface, the process in Fig. 1 was taken. It is obvious that the normal state of (Cu,C)-1234 shows a characteristic of saturation. The emergence of resistance saturation may be derived from the MIR limitation [24]influenced by the carrier density influenced by the oxygen contents or possible underdoping [22] compared with the highest $T_{c0}$ in (Cu,C)-1234 as 96 K[23]. Fig. 3 (c) shows R-T curves of another heterostructure sample under 0-9 T. The $T_{c0}$ under 0T was 82 K. In 9T the sample maintained its superconductivity with $T_{c0}$ above 60 K indicating the good quality.

After the whole process in Fig. 1, the electrical transport properties of YBCO thin film and (Cu,C)-1234 thin films with different thicknesses were carried out to make comparison with the ones prepared solely. Fig. 3 (d) shows the electronic transport properties of YBCO after the deposition of (Cu,C)-1234 and The YBCO thin film got the lower $T_{c0}$ from 91 K to 78 K and the lower $T_c^{onset}$ reduced from 93 K to 87 K meanwhile the (Cu,C)-1234 thin film with thickness of 188 nm got a lower $T_{c0}$ from 80 K to 73.5 K indicated that the YBCO underwent the growth process of (Cu,C)-1234 and (Cu,C)-1234 grown on YBCO layer both lowered the quality than solely prepared ones. For YBCO, the degeneration could be caused by the loss of oxygen content when depositing (Cu,C)-1234 as it underwent another vacant period. For (Cu,C)-1234, the major difference between films solely grown and obtained simultaneously with the heterostructure is the relative position of the growth area and the $CO_2$ nozzle. Even for the same flow rate of $CO_2$, the gradient of the actual concentration near the surface of the substrate could be significantly different. Critically, (Cu,C)-1234 thin films extremely sensitive to the $CO_2$ distribution and concentration, so it is unavoidable that the property more or less variate because of the alternation of the position and the size of growth region. Even though the quality of single films had slightly declined, they still possess nice superconductivity and maintained major features for interface analysis.

Fig. 3 (f) and (g) are thickness-various (Cu,C)-1234 fabricated during the growth process of the heterostructure. The result for 1.2 nm sample was not shown here because of unavoidable test error and the 188 nm sample got fluctuation because of connection problem but show obvious superconductivity transition. Compared with the 188 nm sample, the $T_{c0}$ of 87 nm decreased from 73.5 K to 60.0 K and the resistance of normal state turned out to be approximately double the former. However, when the thickness approached 18 nm, (Cu,C)-1234 lost superconductivity. As the temperature decrease, the resistance inclined to be infinitely large. The resistance decreased and inclined to be stable as the temperature rise to room temperature. These indicates that as the thickness decrease, (Cu,C)-1234 thin films experience the transition from superconductor to insulator.

### 3.3 Superconductivity recovery in (Cu,C)-1234/YBCO heterostructures

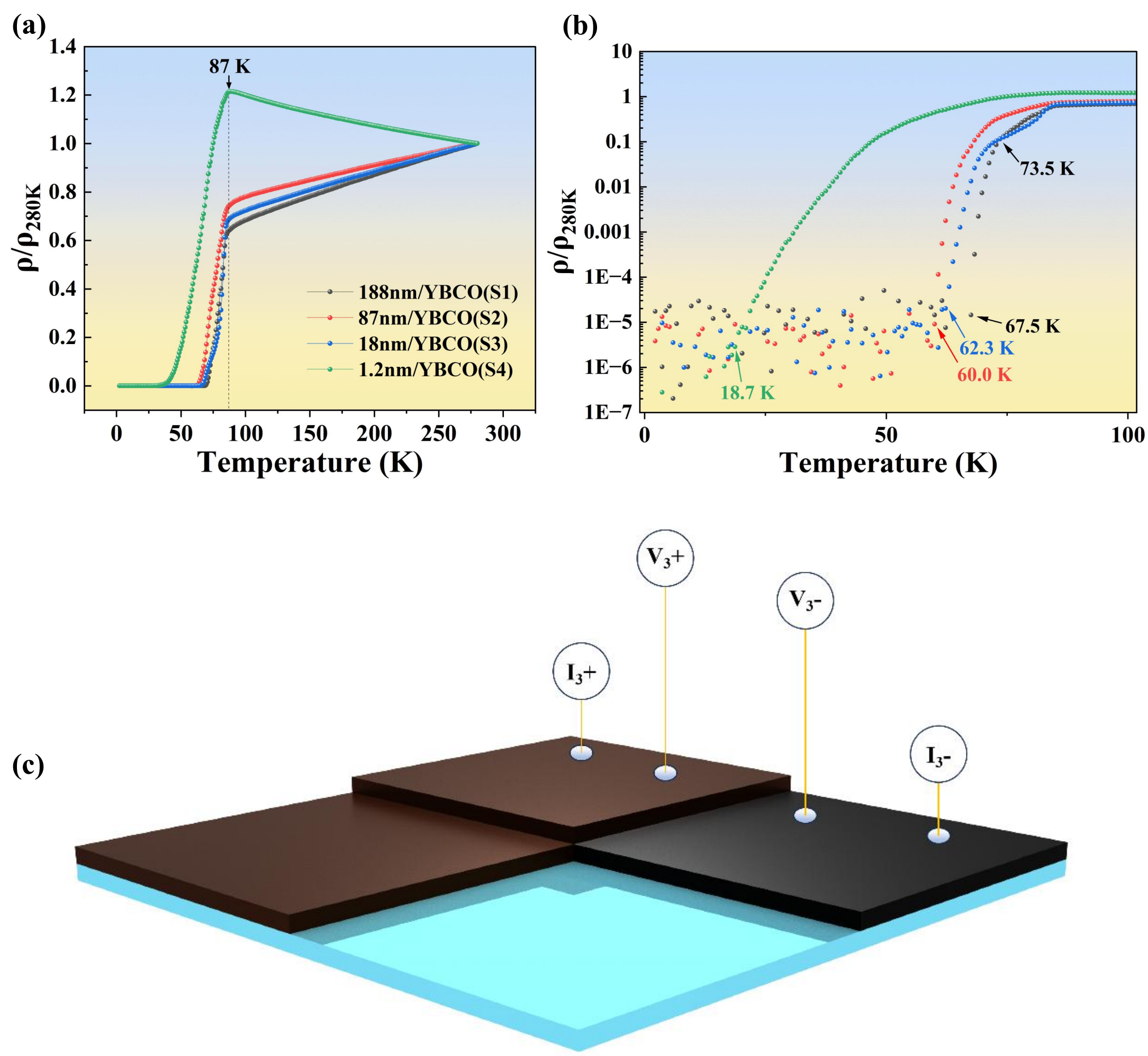

**Fig. 4.** (a) Normalized ρ-T curves of the circuit going through the interface with different thickness of top (Cu,C)-1234 layer controlled by deposition duration. S1-S4 respectively correspond to the thickness of (Cu,C)-1234 as 188 nm, 87 nm, 18 nm and 1.2 nm. (b) Enlarged view of region near $T_{c0}$ in (a). (c)Test circuit for the interface.

The electrical transport properties of circuit through the interface of (Cu,C)-1234/YBCO heterostructures with different thickness of (Cu,C)-1234 were shown in Fig. 4. The thickness of bottom YBCO layer were fixed at approximately 150 nm. Top (Cu,C)-1234 with deposition duration from 4 s to 10 min as the variation and the actual thicknesses were 188 nm , 87 nm , 18 nm and nearly 1.2 nm (estimated by others) for S1-S4, respectively. As the thickness of top layer decrease, S1-S3 got minor performance change in $T_{c0}$ (67.5 K, 60.0 K and 62.3 K) and nearly the same $T_c^{onset}$ as YBCO in Fig. 3 (d). However, as the thickness of (Cu,C)-1234 reached about 1.2 nm, though $T_c^{onset}$ is still 87 K, it underwent a dramatically decrease to 18.7 K meanwhile a linear negative temperature coefficient in normal state. In, Fig. 3 (g), (Cu,C)-1234 thin films below 18 nm have already become insulator, but S3 and S4 both exhibited superconductivity which indicated that the circuit going through the interface is superconducting which suggests the recovery of superconductivity in (Cu,C)-1234.

## 4. Discussion and Conclusion

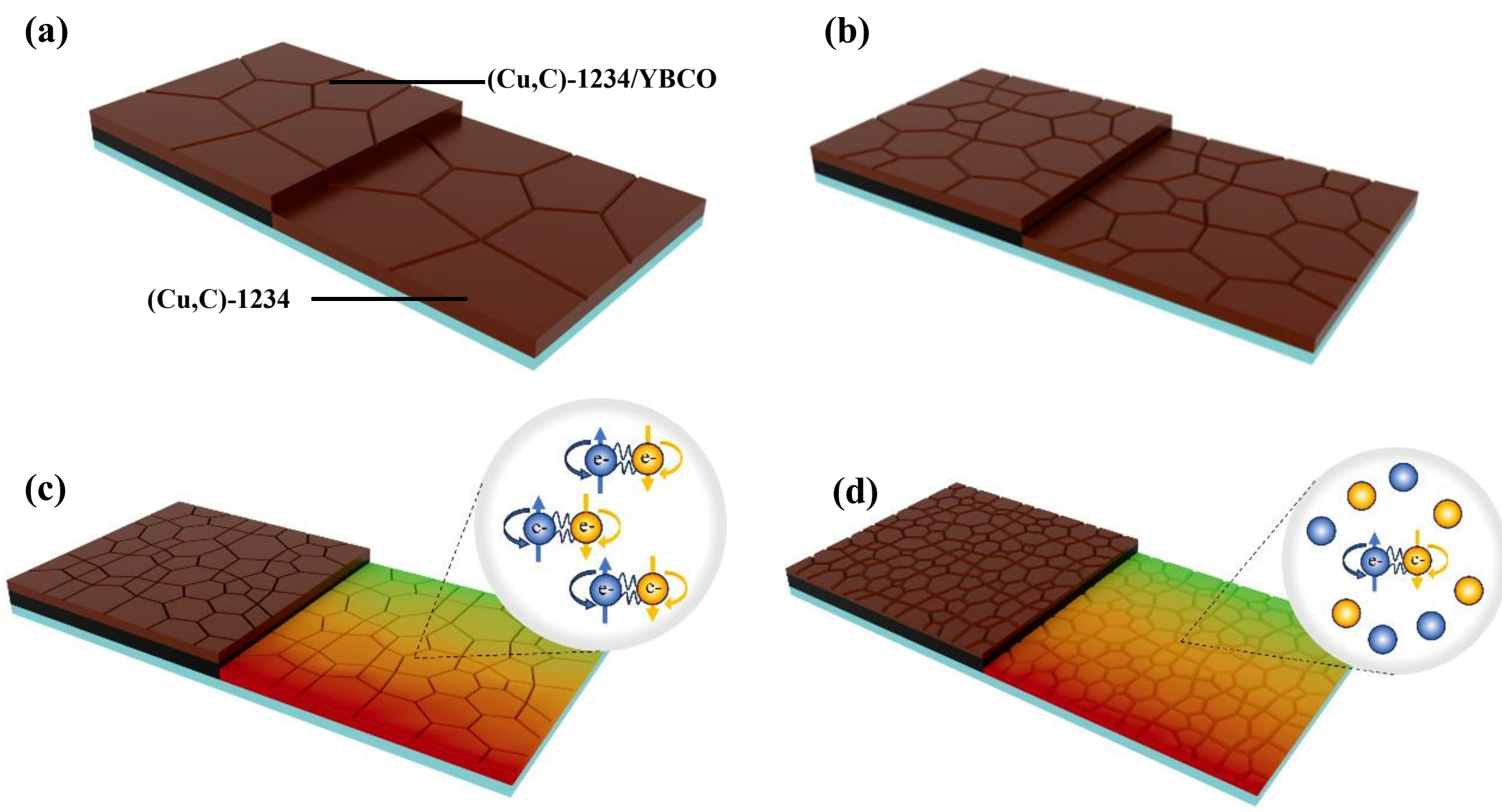

**Fig. 5.** Property evolution of (Cu,C)-1234 and heterostructure with the thickness of (Cu,C)-1234 as (a) 188 nm, (b) 88 nm, (c)18 nm and (d) 1.2 nm

The transport measurements establish two key experimental observations: a thickness-driven SIT in isolated (Cu,C)-1234 and the recovery of superconductivity in (Cu,C)-1234/YBCO heterostructure. Together with the structural characterization, these results allow a physical picture of the superconducting evolution to be constructed, as schematically illustrated in Fig. 5.

For isolated (Cu,C)-1234 shown in Fig. 3 (a) and (f), reducing the thickness progressively suppresses the superconducting transition temperature and increases the normal-state resistance. The saturation in normal state of (Cu,C)-1234 suggests the normal state resistance may approach the MIR limit [24]. Ioffe-Regel criterion [25]claims that the notion of quasiparticle breaks down when the mean free path *l* becomes comparable to the lattice constant a and thus the material approaches the minimal metallic conductivity proposed by Mott[26]. For the thickest 188 nm sample, the relatively low $T_{c0}$ [23]and saturated resistance indicated that the (Cu,C)-1234 itself has already possessed certain disorder. As the thickness decreases, $T_{c0}$ decreases from 73.5 K to 60.0 K in curve of 87 nm sample. When to 18 nm, (Cu,C)-1234 became totally an insulator with infinite resistance to low temperature and the resistance decreased exponentially as the temperature rose. In (Cu,C)-1234, decreased-thickness-induced enhancement of disorder gradually broke down the superconductivity and finally led to its vanishment turning out to be the insulating film. It is believed that the existence of Cooper pairs and their long-range coherence are necessary to realize superconductivity [27-28]. And the disorder could be harmful to both of them[29-32].

The heterostructures exhibit a markedly different evolution. In Fig. 4 (a), the $T_{c0}$ of interface with 188 nm and 87 nm (Cu,C)-1234 both got little difference with isolated (Cu,C)-1234 with the same thickness. The normal state with linear increase as the temperature rose is consistent with YBCO as the strange metal and the $T_c^{onset}$ is the same as YBCO after the growth process of (Cu,C)-1234. In this situation the two layers are largely independent and the curves turned out to be simply the superposition of YBCO and (Cu,C)-1234 layers. However, for the heterostructure with 18 nm (S3) and 1.2 nm (Cu,C)-1234 (S4), things became different. When below 18 nm, (Cu,C)-1234 thin film should have transited into an insulator. But for S3 the superconductivity recovered and the curve of S3 is similar to S1 and S2 which indicated that the superconductivity in (Cu,C)-1234 totally recovered because of the induction of proximity effect between (Cu,C)-1234 and YBCO. For S4, it is worth noted that 1.2 nm is a special scale which is comparable to coherent length of cuprates superconductors. Compared with S1-S3, the $T_{c0}$ of S4 got a dramatically decrease to 18.7 K and in normal state the resistance decreased linearly as the temperature rose. The recovery of superconductivity in S3 is stronger than S4 which is inconsistent with

From 188 nm to 87 nm, the $T_{c0}$ of (Cu,C)-1234 dropped from 73.5 K to 60 K and the $T_{c0}$ of the interface circuit dropped from 67.5 K to 60 K, which indicates the enhancement of disorder. When the thickness approached 18 nm, the disorder is strong enough to localize Cooper pairs and break long range coherence of them in (Cu,C)-1234 thus the SIT in intrinsic superconductor (Cu,C)-1234 emerged. Although the isolated film has already become insulating, the corresponding heterostructure exhibits a zero-resistance superconducting state with a transition temperature comparable to those of the thicker heterostructures. For S3 the insulating (Cu,C)-1234 part restored its superconductivity with $T_{c0}$ as 62.3 K which is close to the $T_{c0}$ of S1 and S2 with superconducting (Cu,C)-1234 (67.5 K and 60.0 K) top layer. In contrast to pure colors, the color gradient in Fig.5 (c) and (d) represent the loss of concordant phase of the Cooper pairs in (Cu,C)-1234. And as the thickness finally decreased to 1.2 nm in S4, the emergence of negative temperature coefficient and the dramatic decrease of $T_{c0}$ simultaneously suggest further process of SIT inside (Cu,C)-1234 meanwhile the improved level of disorder and wave function localization. Not only phase coherence, but also the coupling of Cooper pairs has been destroyed to some extent so that proximity-induced restored superconductivity in S4 was weakened.

Different from S-N or S-F proximity systems, insulating 18 nm (Cu,C)-1234 is an intrinsic superconductor with Cooper pairs localized without long range coherence which made it insulating in ultrathin condition. The non-monotonic minor change of $T_{c0}$ of S1-S3 supports that the recovery of superconductivity was not simply induced by injection of Cooper pairs in YBCO since in this way the $T_{c0}$ should be closer to 78 K as $T_{c0}$ of YBCO and the thinner (Cu,C)-1234 is the easier injection happens which is inconsistent with the result in S4 whose superconductivity significantly degraded and insulating characteristic enhanced leading to negative temperature coefficient. The main role here YBCO displayed is that it reestablished the phase coherence inside ultrathin (Cu,C)-1234 layer. Actually, for heterostructure with ultrathin (Cu,C)-1234 the intrinsic superconductivity unlocking via phase reference took place and as the thickness decreased the disorder together with localization continued to strengthen which damaged superconducting properties firstly the phase coherence. When got extremely thin like 1.2 nm which is comparable to coherent length here, the pairing of Cooper pairs inside (Cu,C)-1234 also degraded which led to much lower $T_{c0}$ and stronger insulating property in heterostructure. In this way, robust superconductor YBCO provided not simply Cooper pairs but phase coherence reference [29] to help localized Cooper pairs of (Cu,C)-1234 to rebuild long range coherence so that the superconductivity of (Cu,C)-1234 recovered.

To explore more detailed microscopic process, further verifications focused on the evolution of electrical structure and energy band are worth carrying out to (Cu,C)-1234/YBCO and similar S-S' heterostructure as a perspective to reveal unclear superconducting microscopic mechanism.

In summary, epitaxial (Cu,C)-1234/YBCO heterostructures were successfully fabricated to investigate the interplay between the superconducting proximity effect and the thickness-driven superconductor–insulator transition (SIT). Isolated (Cu,C)-1234 films exhibit a thickness-driven SIT at a critical thickness of approximately 18 nm. In contrast, superconductivity is re-established in transport measurements when the insulating (Cu,C)-1234 layer is coupled to superconducting YBCO, demonstrating a robust superconducting proximity effect. As the (Cu,C)-1234 thickness is further reduced to approximately 1.2 nm, the recovered superconductivity is markedly suppressed, indicating that besides the loss of phase coherence the intrinsic superconductivity degraded.

The observed behavior is consistent with a picture in which interfacial coupling restores superconductivity in ultrathin (Cu,C)-1234 after the thickness-driven SIT, while increasing disorder and dimensional confinement progressively suppress this effect as the film approaches the thickness comparable to coherent length. These findings demonstrate that interface engineering provides an effective route to manipulating superconductivity in ultrathin cuprates and establish (Cu,C)-1234/YBCO heterostructures as a useful platform for investigating the interplay between dimensionality, superconducting proximity effects, and quantum phase transitions.

**Data availability statement**

Data available upon request.

**Conflict of interest**

The authors declare that they have no conflict of interest.

**Acknowledgments**

This work was supported by Shanghai Science and Technology Innovation Program (Grant No. 23511101600), National Science Foundation of China (Grant Nos. 52477022, 52172271, 12374378, 52307026), the National Key R&D Program of China (Grant No. 2022YFE03150200), Ministry of Industry of Information Technology project (Grant No. TC230H0AC/159).

**Author Contributions**

Zhihang Xu

Data curation (equal), Formal analysis (equal), Investigation (equal), Writing-original draft (equal)

Detian Yang

Investigation (equal), Analysis (equal), Consultation (equal)

Ping Zhu

Investigation (equal)

Ruoxian Sun

Investigation (equal)

Xiaoyang Cai
Investigation (equal)

Yanqun Guo
Supervision (equal), Writing-review & editing (equal)

Chuanbing Cai
Supervision (equal), Writing-review & editing (equal)